%% file: main.tex
\documentclass[10pt,conference]{IEEEtran}
\IEEEoverridecommandlockouts
\usepackage[top=0.7in, bottom=1.0in, left=0.625in, right=0.625in]{geometry} 
\usepackage{cite}
\usepackage{amsmath,amssymb,amsfonts}
\usepackage{algorithm}
\usepackage{algorithmic}
\usepackage{graphicx}
\usepackage{textcomp}
\usepackage{xcolor}
\usepackage{siunitx}
\usepackage{booktabs}
\usepackage{multirow}
\usepackage{subfig}
\usepackage[font=small]{caption}

\def\BibTeX{{\rm B\kern-.05em{\sc i\kern-.025em b}\kern-.08em
    T\kern-.1667em\lower.7ex\hbox{E}\kern-.125emX}}

\usepackage{fancyhdr}
\fancypagestyle{firststyle}{\fancyhf{}

		\fancyhead[L]{ \small  M. Ying, P. Ma, D. Shakya, T. S. Rappaport, ``Multi-Stage Location Optimization Through Power Delay Profile Alignment Using Site-Specific Wireless Ray Tracing," in \textit{GLOBECOM 2025 - 2025 IEEE Global Communications Conference}, Taipei, Taiwan, Dec. 2025, pp. 1-6.}
	}
\usepackage{times}

\begin{document}
\bstctlcite{IEEEtran:BSTcontrol}
\title{Multi-Stage Location Optimization Through Power Delay Profile Alignment Using Site-Specific Wireless Ray Tracing}

\author{\IEEEauthorblockN{Mingjun Ying, Peijie Ma, Dipankar Shakya, and Theodore S. Rappaport}
\IEEEauthorblockA{\textit{NYU WIRELESS, Tandon School of Engineering, New York University, USA}\\
    \{yingmingjun,
     pm3688, dshakya, 
     tsr\}@nyu.edu}
\thanks{This research is supported by the New York University (NYU) WIRELESS Industrial Affiliates Program and NSF award 2234123.}}

\maketitle

	\thispagestyle{firststyle}

\begin{abstract}
    Ray tracing (RT) simulations require accurate transmitter (TX) and receiver (RX) location information from real-world measurements to accurately characterize wireless propagation behavior in an environment. Such wireless propagation measurements typically employ GPS-based logging for TX/RX locations, which can produce meter-level errors that lead to unreliable RT calibration and validation. These location misalignments cause inaccurate interactions between RT-generated multipath components (MPCs) and the modeled 3D environment, which lead to erroneous channel predictions, and severe discrepancies between simulated and measured power delay profiles (PDPs) and channel characteristics. Moreover, the same RT-generated PDPs using inaccurate locations result in calibration errors when adjusting material properties such as conductivity and permittivity. 
    This paper presents a systematic multi-stage TX/RX location calibration framework to correct location errors and consequently align measured and simulated omnidirectional PDPs. 
    Optimization is performed using a computationally efficient multi-stage grid search and the Powell method. Applying the location calibration framework to NYU WIRELESS urban-microcell (UMi) measurements at 6.75 GHz and 16.95 GHz corrected TX/RX location errors of up to 7 m. The framework reduced the composite loss function by 42.3\% for line-of-sight (LOS) and 13.5\% for non-line-of-sight (NLOS) scenarios. Furthermore, peak power prediction accuracy improved by approximately 1 dB on average.  Such improved geometric alignment enables accurate channel prediction, vital for beam management and infrastructure deployment for next-generation wireless networks.
\end{abstract}


\input{sections/introduction}

\input{sections/nyuray_overview}
\input{sections/environment_modeling}
\input{sections/location_calibration}
\input{sections/results}

\input{sections/conclusion}


\bibliographystyle{IEEEtran}
\bibliography{references}

\end{document}

%% file: sections/introduction.tex
\section{Introduction}
\label{sec:introduction}
Accurate wireless channel modeling serves as the foundation for effective design, deployment, and optimization of next-generation communication systems~\cite{rappaport2024wirelessbook}.
The channel modeling accuracy is vital for future wireless networks adopting upper mid-band and mmWave frequencies~\cite{rappaport2019ia, Shakya2024ojcom} and complex technologies like massive MIMO and reconfigurable intelligent surfaces (RIS)~\cite{bjornson2022reconfigurable, basar2019wireless}. Accurate channel predictions enable precise beamforming, coverage planning, and capacity estimation, all crucial for effective beam management and infrastructure deployments~\cite{rappaport2017overview,Shakya2024ojcom}.
Wireless ray tracing (RT) has emerged as the preferred deterministic channel modeling approach due to the ability to explicitly account for physical propagation phenomena in site-specific 3D environments~\cite{mckown1991ray, seidel1994tvt}. Unlike statistical models (e.g., 3GPP TR 38.901~\cite{3GPPTR38901}) that provide average channel behavior, RT simulates electromagnetic (EM) wave interactions with specific environmental objects, capturing reflection, penetration, diffraction, and scattering effects that directly influence the path loss, delay spread, and angular spread of the wireless channel.


Traditional wireless RT calibration primarily focuses on reflection and penetration loss for different materials across frequencies~\cite{hoydis2023sionna,kanhere2024calibration,ruah2024calibrating}, to minimize differences between measured and simulated wireless channels. A critical assumption in traditional calibration approaches is the perfect geometric alignment of the simulation setup with the physical measurement environment. However, a significant limitation stems from uncertainty in the recorded TX and RX locations used as inputs for RT simulations. For instance, measurement campaigns, including recent upper mid-band and past mmWave and sub-THz efforts at New York University (NYU)~\cite{rappaport2013ia, maccartney2015ia, Xing2021_Inicl, Shakya2024ojcom, Ju2024twc}, often relied on consumer-grade Global Positioning System (GPS) for logging TX and RX coordinates, which can introduce location errors of 5-10 meters~\cite{wang2020precision}.  Such positional errors from GPS lead to the mismatches between the actual TX/RX locations during measurement and the coordinates used within the RT simulation, potentially invalidating calibration results that neglect location uncertainty.
Precise TX-RX location alignment is vital for the effective use of site-specific wireless propagation measurement data at different frequencies. 

Despite the importance of TX/RX location calibration for RT validation, systematic methodologies for the calibration are notably underrepresented in published research. Furthermore, popular RT tools like Wireless InSite \cite{RemcomWirelessInSite} and Sionna \cite{hoydis2023sionna} do not provide functionality for addressing location uncertainty. The lack of robust, automated methods for location calibration cripples RT simulation accuracy and reusability of measured data at different frequencies. 
To address the critical need for accurate TX/RX location matching between measurement and RT simulations, this paper introduces a systematic method for location calibration. The novel method aligns measured and RT-simulated omnidirectional PDPs using a multi-component loss function and an optimization algorithm to determine optimal location adjustments. The key contributions of the paper are:

\begin{itemize}
    \item A mathematical formulation of the TX/RX location optimization problem is presented to enable calibration. 
      
    \item A multi-component loss function is introduced to quantify discrepancies between measured and RT-simulated PDPs, considering multipath component (MPC) delay, power, and overall PDP shape.

    \item An efficient multi-stage location optimization algorithm is detailed, combining grid search with local refinement using the Powell method, to determine optimal TX/RX location adjustments.
    
    \item Experimental validation is presented using upper mid-band (6.75 GHz and 16.95 GHz) UMi measurements from NYU WIRELESS~\cite{Shakya2025icc} demonstrating substantial reduction of
    composite loss function by 42.3\% for LOS and 13.5\% for NLOS locations.
\end{itemize}

The remainder of the paper is organized as follows. Section \ref{sec:nyuray_overview} presents NYURay, the deterministic ray‑tracing engine developed at NYU WIRELESS. Section \ref{sec:environment_modeling} outlines the 3D environment modeling method, and Section \ref{sec:location_calibration} details the TX/RX location calibration algorithm. Section \ref{sec:results} validates the calibrated RT predictions against measurement results. Section \ref{sec:conclusion} offers concluding remarks.

%% file: sections/nyuray_overview.tex
\section{Overview of NYURay} 
\label{sec:nyuray_overview}
NYURay is a wireless RT simulation tool developed at NYU WIRELESS for deterministic channel modeling and analyzing radio wave propagation in 3D environments~\cite{kanhere2024calibration}.

NYURay implements four fundamental propagation mechanisms to simulate electromagnetic wave interactions with the environment: (1) Reflection: computed using a hybrid approach combining Shooting and Bouncing Rays
~\cite{kanhere2024calibration}, image-based RT, and Fresnel reflection coefficients calculated based on material properties, polarization, and incidence angle~\cite{ITU-P2040-3}; (2) Penetration: modeled for waves transitioning between media with the overall loss, accounting for both boundary reflections and internal absorption~\cite{ITU-P2040-3}; (3) Diffraction: implemented using Uniform Theory of Diffraction~\cite{mcnamara1990introduction} where rays emanate from edges along Keller cones~\cite{keller1962geometrical} with computed diffraction coefficients dependent on incidence angle, observation angle, and edge geometry; and (4) Scattering: applied for rough surfaces exceeding the Rayleigh criterion~\cite{rappaport2024wirelessbook}, with scattered energy distribution following the Lambertian pattern~\cite{degli2001diffuse}. The combined propagation mechanisms generate a comprehensive multipath component that captures delay, power, and angular characteristics of the wireless channel.

%% file: sections/environment_modeling.tex
\section{Environment Modeling for Ray Tracing}
\label{sec:environment_modeling}
Accurate 3D environment models are crucial for reliable ray tracing simulations and subsequent TX/RX location calibration. High-fidelity models capture the necessary geometric and electromagnetic characteristics of the propagation environment. For outdoor UMi 3D environment modeling, a specific process using OpenStreetMap (OSM) data is implemented.

The UMi modeling workflow involves four steps. First, OSM geographical data, including terrain elevation and initial building structures, are imported using the BLOSM addon for Blender~\cite{blosm} to establish the ground reference plane and basic geometry. Second, the imported building models are manually adjusted using precise measurements from laser rangefinders and iPhone LiDAR scans to match actual dimensions. Furthermore, numerous manually created objects, such as trees, lamp posts, traffic signs, benches, and fire hydrants, are incorporated to enhance the fidelity of the 3D model and accurately represent the propagation environment. Third, material properties (e.g., concrete, glass, metal, wood) are assigned to the surface of each object, corresponding to frequency-dependent EM characteristics utilized by the NYURay simulator~\cite{ITU-P2040-3}. Fourth, the final 3D model is exported into Mitsuba XML format for compatibility with NYURay EM field calculations.

\begin{figure}[!t]
    \centering
    \includegraphics[width=0.98\columnwidth]{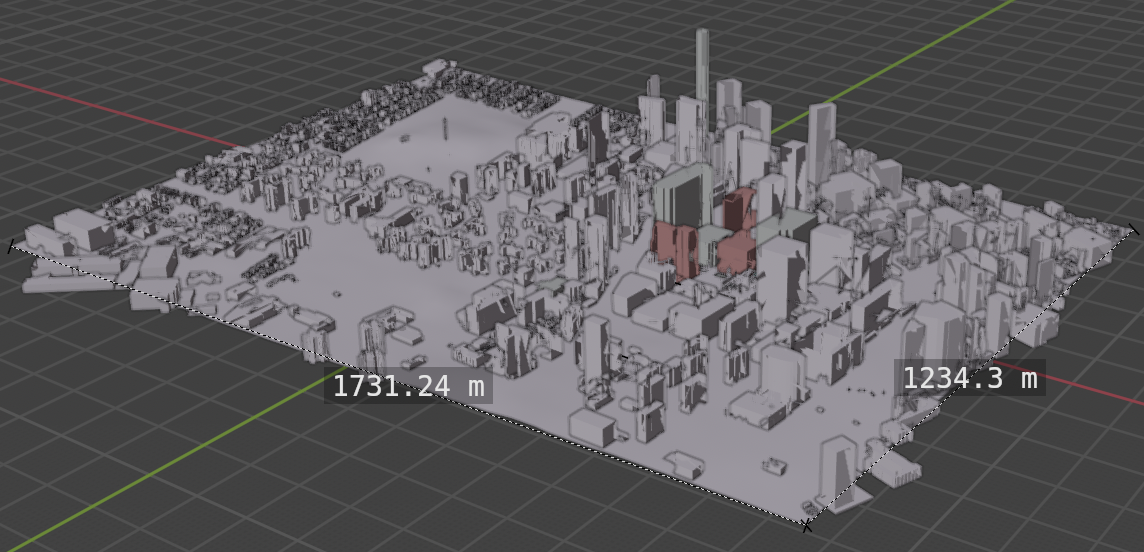}
    \caption{3D visualization of the UMi environment model used in NYURay simulations, covering an area of approximately \SI{1.2}{\kilo\meter} $\times$ \SI{1.7}{\kilo\meter}. Each gray grid represents a 100\,m $\times$ 100\,m area.
     }
    \label{fig:environmental_details}
    \vspace{-15pt}
\end{figure}

Fig.~\ref{fig:environmental_details} illustrates the isometric view of the detailed UMi 3D model. 
Precise geometry (building dimensions, object placement) and correct material properties minimize the need for the location calibration algorithm to compensate for modeling inaccuracies. Detailed environmental features ensure simulated multipath components closely match real-world measurements, leading to more reliable calibration results.

%% file: sections/location_calibration.tex
\section{TX-RX Location Calibration Method}
\label{sec:location_calibration}

This section presents a systematic TX-RX location optimization method by minimizing discrepancies between simulated and measured PDPs. The proposed location calibration method addresses GPS-induced location uncertainty in field measurements, enhancing RT validation accuracy through systematic location refinement.

\subsection{Problem Formulation}

Initial TX and RX locations, $\mathbf{p}_{TX}^{(0)}$ and $\mathbf{p}_{RX}^{(0)}$, are derived from GPS coordinates recorded during the measurement campaign. These geographic coordinates (latitude, longitude, typically World Geodetic System WGS84 standard) are transformed into a local Cartesian coordinate system (x, y, z) aligned with the 3D environment model used for ray tracing. Coordinate conversion utilizes an Azimuthal Equidistant (AEQD) projection centered on the measurement area to yield local (x, y) values in meters~\cite{snyder1997flattening}. Initial z-coordinates for TX and RX are set based on antenna heights above the local terrain, specifically 4 m for the TX and 1.5 m for the RX.
\begin{equation}
\mathbf{p}_{TX}^{(0)} = [x_{TX}^{(0)}, y_{TX}^{(0)}, z_{TX}^{(0)}], \,\,
\mathbf{p}_{RX}^{(0)} = [x_{RX}^{(0)}, y_{RX}^{(0)}, z_{RX}^{(0)}].
\end{equation}

The objective is to find location adjustments $\Delta\mathbf{p}_{TX}$ and $\Delta\mathbf{p}_{RX}$ such that:
\begin{equation}
\mathbf{p}_{TX}^{*} = \mathbf{p}_{TX}^{(0)} + \Delta\mathbf{p}_{TX}, \,\,
\mathbf{p}_{RX}^{*} = \mathbf{p}_{RX}^{(0)} + \Delta\mathbf{p}_{RX}.
\end{equation}

The adjusted locations should minimize the discrepancy between simulated and measured PDPs:
\begin{equation}
\{\Delta\mathbf{p}_{TX}^*, \Delta\mathbf{p}_{RX}^*\} = \arg\min_{\Delta\mathbf{p}_{TX}, \Delta\mathbf{p}_{RX}} L(P_{sim}, P_{meas}),
\end{equation}
where $L(\cdot,\cdot)$ represents a loss function quantifying the differences between the simulated omnidirectional PDP $P_{sim}$ (generated using the adjusted locations) and the synthesized omnidirectional PDP $P_{meas}$ from measurements~\cite{Shakya2025icc,sun2015gc}.

To maintain physical plausibility and limit the search space, constraints are imposed:
{\small
\begin{equation}
\|\Delta\mathbf{p}_{TX}\| \leq d_{\max},\,\, \|\Delta\mathbf{p}_{RX}\| \leq d_{\max},\,\,
\Delta z_{TX} = \Delta z_{RX} = 0,
\end{equation}}
where $d_{\max}$ represents the maximum allowable location adjustment (typically 10 meters based on practical GPS accuracy limitations~\cite{wang2020precision}). The height constraint is imposed because the vertical antenna positions are measured with laser rangefinders during field measurements.

\subsection{Ray Tracing Simulation and PDP Generation}

The simulation of PDPs employs NYURay, which models EM wave propagation including direct LOS paths, reflections from surfaces, and diffractions around edges. For each TX-RX pair, the simulator computes paths with associated complex amplitudes $a_i$ and delays $\tau_i$. The channel impulse response (CIR) is formulated as:
\begin{equation}
    h(t) = \sum_{i=1}^{N_p} a_i \delta(t - \tau_i),
\end{equation}
where $N_p$ is the number of paths and $\delta(\cdot)$ is the Dirac delta function. The PDP is then calculated as the squared magnitude of the CIR:
\begin{equation}
    P(\tau) = |h(t)|^2 = \sum_{i=1}^{N_p} |a_i|^2 \delta(t - \tau_i).
\end{equation}

The measured PDP is typically averaged (e.g., over time or local space) to average out small-scale fading effects, yielding an estimate of the local average PDP. Comparing the simulated PDP against this averaged measured PDP is appropriate for location calibration, as the objective is to match MPCs whose delays and powers are primarily determined by the TX/RX locations and surrounding environment structures.

\subsection{PDP Alignment Techniques}

Before comparing PDPs, proper temporal alignment must be established to account for different reference points and timing offsets. Two alignment strategies are employed:

\subsubsection{Maximum Peak Alignment}
The most straightforward approach aligns PDPs based on their maximum peaks, typically corresponding to the dominant path (often LOS when present):
\begin{equation}
\Delta\tau_{simple} = \tau_{sim}^{\max} - \tau_{meas}^{\max},
\end{equation}
where $\tau_{sim}^{\max}$ and $\tau_{meas}^{\max}$ represent the delays corresponding to the maximum power in the simulated and measured PDPs, respectively. This alignment works reliably when a dominant path exists for both RT simulation and measurement.

\subsubsection{Correlation-Based Multi-Peak Alignment}
In complex multipath environments where multiple significant peaks exist (e.g., NLOS scenarios or LOS within narrow street canyons), a more robust approach finds the optimal delay shift maximizing correlation:
\begin{equation}
\Delta\tau_{multi} = \arg\max_{\Delta\tau} \,\,\text{corr}(P_{sim}(\tau), P_{meas}(\tau + \Delta\tau)).
\end{equation}
The alignment method selection depends on the achieved correlation value; multi-peak alignment is preferred if it produces a correlation coefficient exceeding 0.5, otherwise, the maximum peak alignment is used.

\vspace{-5pt}

\subsection{Multi-Component Loss Function}
The loss function combines multiple specialized components to comprehensively evaluate PDP similarity from different perspectives:

\subsubsection{Peak Matching Loss}
The peak matching component focuses on aligning significant MPCs, quantifying both temporal and power discrepancies:
{\small
\begin{equation}
\begin{split}
L_{peak} = \frac{1}{N_{matched}} \sum_{i=1}^{N_{sim}} \min_{j} & \biggl[ w_i   \biggl( \frac{|\tau_{sim,i} - \tau_{meas,j}|}{T_{norm}} \\
& + |P_{sim,i}' - P_{meas,j}'| \biggr) \biggr],
\end{split}
\end{equation}}
where $N_{matched}$ is the number of matched peaks; $w_i = e^{-\tau_{sim,i}/\tau_{ref}}$ is an exponential weight prioritizing earlier simulated paths, with the reference delay $\tau_{ref}$ set to 500 ns (corresponding to approximately 150 m path length, typical for early MPCs in a UMi environment). The early arriving MPCs contribute more significantly to overall channel capacity and are more sensitive to location changes. $\tau_{sim,i}$ and $\tau_{meas,j}$ represent the arrival times of the $i$-th simulated and $j$-th measured peaks, respectively; $T_{norm} = 100$ ns is a normalization factor scaling delay differences to be comparable with power differences; and $P_{sim,i}'$, $P_{meas,j}'$ are the corresponding peak powers normalized to a [0, 1] scale.

\subsubsection{Unmatched Peaks Penalty}
The unmatched peaks component encourages simulations that produce similar MPC richness to measurements:
\begin{equation}
L_{unmatched} = w_{unmatched} \cdot \frac{|N_{sim} - N_{meas}|}{\max(N_{sim}, N_{meas})},
\end{equation}
where $N_{sim}$ and $N_{meas}$ represent the number of significant peaks\footnote{A peak is considered significant when it meets three criteria: (1) its power is within a 25 dB threshold below the maximum power in the PDP, (2) it has sufficient prominence (at least 5\% of the threshold value or above 1 dB, whichever is larger) to stand out from surrounding delay bins, and (3) it is separated from other peaks by a minimum distance (3 bins) in the delay domain. Additionally, peaks are prioritized by power and filtered to ensure a minimum temporal separation (15 ns), preserving only distinct MPCs.} in simulated and measured PDPs, and $w_{unmatched} = 0.5$ is a weighting factor balancing with peak matching. The normalization by the maximum peak count produces a relative measure (0-1) of multipath count mismatch.

\subsubsection{Shape Loss}
The shape loss captures the overall PDP distribution similarity beyond just discrete peaks:
\begin{equation}
L_{shape} = \frac{1}{|\mathcal{R}|} \sum_{\tau \in \mathcal{R}} (P_{sim,dB}(\tau) - P_{meas,dB}(\tau))^2,
\end{equation}
where $P_{dB}$ represents power values in dB scale, normalized to 0 dB maximum, $\mathcal{R}$ denotes the set of delay bins where either PDP exceeds the significance threshold (25 dB below the peak), and $|\mathcal{R}|$ is the cardinality of considered delay bins. Shape loss employs mean squared error in dB scale to evaluate continuous similarity across all delay values.

The final loss combines all components with regularization:
\begin{equation}\label{eq:loss_function}
L = \alpha \cdot (L_{peak} + L_{unmatched}) + (1-\alpha) \cdot L_{shape} + \beta \cdot L_{distance},
\end{equation}
where $\alpha=0.7$ weights the peak-based metrics relative to the shape metric, $L_{distance} = (\|\Delta\mathbf{p}_{TX}\|^2 + \|\Delta\mathbf{p}_{RX}\|^2)/(d_{\max}^2)$ is a regularization component penalizing large TX/RX location adjustments from initial GPS coordinates, and $\beta=0.05$ controls the regularization influence. 

\subsection{TX-RX Location Optimization Algorithm}

The optimization framework combines alternating minimization with hierarchical grid refinement and gradient-free optimization, balancing computational efficiency with solution accuracy for the complex, non-convex loss landscape.

\subsubsection{Alternating Minimization Strategy}

The algorithm decomposes the joint 4D optimization problem into sequential 2D subproblems through alternating minimization. This approach iteratively optimizes:
\begin{align}
\mathbf{p}_{RX}^{(k+1)} &= \arg\min_{\mathbf{p}_{RX}} L(\mathbf{p}_{TX}^{(k)}, \mathbf{p}_{RX}) \\
\mathbf{p}_{TX}^{(k+1)} &= \arg\min_{\mathbf{p}_{TX}} L(\mathbf{p}_{TX}, \mathbf{p}_{RX}^{(k+1)})
\end{align}

This decomposition reduces computational complexity from $O(N_{grid}^4)$ for simultaneous 4D search to $O(N_{grid}^2)$ for sequential 2D searches—a 300-fold reduction in RT simulations for typical grid resolutions.

\subsubsection{Three-Stage Hierarchical Optimization}

Each 2D subproblem employs a three-stage hierarchical approach that progressively refines the location estimate:

\textbf{Stage 1 - Coarse Grid Search:} 
The initial exploration evaluates a $5 \times 5$ coarse grid spanning $[-5, 5]$ m with 2.5 m spacing. For each candidate position $\mathbf{p}^{cand}$, the algorithm verifies constraints $\|\mathbf{p}^{cand} - \mathbf{p}^{(0)}\| \leq d_{\max}$, executes RT simulation, computes PDP alignment, and evaluates loss function $L^{cand}$. This identifies the global minimum basin while providing robustness against local minima.

\textbf{Stage 2 - Fine Grid Refinement:} 
Centered on the coarse optimum $\mathbf{p}^{coarse}$, fine refinement explores a $\pm$1.5 m region with 0.5 m spacing ($7 \times 7$ grid). This achieves sub-meter accuracy with resolution based on Nyquist sampling for typical urban multipath correlation distances (~1 m).

\textbf{Stage 3 - Gradient-Free Local Optimization:} 
Powell's conjugate direction method~\cite{powell1965method} performs final refinement from $\mathbf{p}^{fine}$, constructing conjugate search directions $\{\mathbf{u}_k\}$ without gradient computation—crucial for non-differentiable RT simulations. 

Starting with coordinate-aligned directions $\mathbf{u}_1 = [1, 0]^T$, $\mathbf{u}_2 = [0, 1]^T$, line minimization determines optimal step sizes:
\begin{equation}
\alpha_{k}^{*} = \arg\min_{\alpha}\, L\bigl(\mathbf{p}^{current} + \alpha \,\mathbf{u}_{k}\bigr),
\end{equation}
subject to $\|\mathbf{p}^{current} + \alpha \,\mathbf{u}_{k} - \mathbf{p}^{(0)}\| \leq d_{\max}$.

Search directions update to maintain conjugacy by replacing maximum improvement directions with $\mathbf{u}_{new} = \mathbf{p}^{end} - \mathbf{p}^{start}$.

\subsubsection{Convergence Criteria and Computational Efficiency}

Optimization terminates when position updates satisfy:
\begin{equation}
\|\mathbf{p}_{TX/RX}^{(k+1)} - \mathbf{p}_{TX/RX}^{(k)}\| < 0.1 \text{ m}
\end{equation}
and relative loss change $|L^{(k+1)} - L^{(k)}|/L^{(k)} < 10^{-3}$. 

The algorithm converges in 3--5 iterations with 500--750 RT simulations total, requiring approximately 3--4 minutes per TX-RX pair on Apple M4 Max (16 cores, 4.51 GHz), dominated by RT simulation time (approximately 0.3 seconds each). 

Algorithm~\ref{alg:txrx_opt_refined} details the complete optimization procedure, with alternating minimization continuing until both TX and RX position updates fall below the threshold $\epsilon = 0.1$ m.

\begin{algorithm}[!t]
\scalebox{0.95}{
\begin{minipage}{\linewidth}
\small
\caption{TX-RX Location Calibration Algorithm}
\label{alg:txrx_opt_refined}
\begin{algorithmic}[1]
\REQUIRE Initial positions $\mathbf{p}_{TX}^{(0)}, \mathbf{p}_{RX}^{(0)}$, measured PDP $P_{meas}$, 3D environment model, NYURay RT engine parameters
\ENSURE Calibrated positions $\mathbf{p}_{TX}^{*}, \mathbf{p}_{RX}^{*}$
\STATE \textbf{Initialize:} $\mathbf{p}_{TX} \gets \mathbf{p}_{TX}^{(0)}$, $\mathbf{p}_{RX} \gets \mathbf{p}_{RX}^{(0)}$
\STATE \textbf{Parameters:} $d_{max} = 10$ m, $\epsilon = 0.1$ m, $N_{max} = 10$
\FOR{$iter = 1$ \textbf{to} $N_{max}$}
    \STATE $\mathbf{p}_{RX}^{prev} \gets \mathbf{p}_{RX}$
    \STATE \textbf{Stage 1:} Coarse grid search for RX
    \FOR{$\Delta x \in \{-5, -2.5, 0, 2.5, 5\}$ m}
        \FOR{$\Delta y \in \{-5, -2.5, 0, 2.5, 5\}$ m}
            \STATE $\mathbf{p}_{RX}^{cand} \gets \mathbf{p}_{RX}^{(0)} + [\Delta x, \Delta y, 0]^T$
            \IF{$\|\mathbf{p}_{RX}^{cand} - \mathbf{p}_{RX}^{(0)}\| \leq d_{max}$}
                \STATE Run RT: $P_{sim} \gets \text{NYURay}(\mathbf{p}_{TX}, \mathbf{p}_{RX}^{cand})$
                \STATE Evaluate: $L_{cand} \gets L(P_{sim}, P_{meas})$ using \eqref{eq:loss_function}
            \ENDIF
        \ENDFOR
    \ENDFOR
    \STATE $\mathbf{p}_{RX}^{coarse} \gets \arg\min L_{cand}$
    \STATE \textbf{Stage 2:} Fine grid refinement around $\mathbf{p}_{RX}^{coarse}$
    \STATE Search $\pm$1.5 m with 0.5 m spacing $\rightarrow \mathbf{p}_{RX}^{fine}$
    \STATE \textbf{Stage 3:} Powell optimization from $\mathbf{p}_{RX}^{fine}$
    \STATE $\mathbf{p}_{RX} \gets \text{Powell}(L, \mathbf{p}_{RX}^{fine}, \text{tol}=0.1)$
    \STATE $\mathbf{p}_{TX}^{prev} \gets \mathbf{p}_{TX}$
    \STATE Repeat Stages 1--3 for TX optimization $\rightarrow \mathbf{p}_{TX}$
    \IF{$\|\mathbf{p}_{TX} - \mathbf{p}_{TX}^{prev}\| < \epsilon$ \AND $\|\mathbf{p}_{RX} - \mathbf{p}_{RX}^{prev}\| < \epsilon$}
        \STATE \textbf{break}
    \ENDIF
\ENDFOR
\RETURN $\mathbf{p}_{TX}^{*} = \mathbf{p}_{TX}$, $\mathbf{p}_{RX}^{*} = \mathbf{p}_{RX}$
\end{algorithmic}
\end{minipage}
}
\end{algorithm}

The calibration framework effectively corrects location errors up to 7 meters introduced by consumer-grade GPS measurements. Such improved geometric alignment enables accurate channel prediction, vital for beam management, infrastructure deployment, and material property calibration across frequencies in next-generation wireless networks. 


%% file: sections/results.tex
\section{Experimental Results}
\label{sec:results}
This section presents experimental validation of the proposed TX-RX location calibration method in the UMi environment shown in Fig. \ref{fig:environmental_details} at upper mid-band frequencies. 

\subsection{NYU UMi Measurements and RT Simulations}

An extensive UMi measurement campaign was conducted at the NYU Tandon School of Engineering campus in Brooklyn, USA. Measurements covered T-R location pairs within the $\sim$180 m open-square MetroTech Commons and along a 1 km path on Myrtle Avenue, with TX and RX positions indicated in Fig. 1 in \cite{Shakya2025icc}.
A total of 20 locations were measured with 7 LOS and 13 NLOS locations and T-R distances ranging from 40 to 1000 m. Two outages at NLOS locations were observed at both 6.75 and 16.95 GHz. Measurement data for RT location calibration and validation were collected using a sliding correlation-based wideband channel sounder operating at 6.75 GHz and 16.95 GHz~\cite{Shakya2024ojcom}. The sounder featured a 1 GHz RF bandwidth and a maximum measurable delay of 4094 ns. Omni PDPs were synthesized using a power threshold of the maximum between 5 dB above the noise floor and 25 dB below the peak power, and spatial combining based on antenna half-power beamwidth~\cite{sun2015gc}.

NYURay RT simulations were configured with isotropic TX/RX antenna patterns (0 dBi gain), 0 dBm transmit power, and carrier frequencies of 6.75 GHz and 16.95 GHz. The simulation parameters included a minimum received power threshold of -160 dBm for PDP cutoff, with reflection and diffraction mechanisms enabled. Noise and bandwidth constraints were omitted assuming idealized channel impulse responses. Initial TX/RX positions were obtained from GPS recordings during the UMi measurement campaign~\cite{Shakya2025icc}, and simulations utilized the detailed 3D environment model depicted in Fig.~\ref{fig:environmental_details}.

\subsection{Calibration Performance}
Table \ref{tab:calibration_metrics} presents key performance metrics for the location calibration algorithm. Across all measured T-R pairs, the calibration yielded an average loss function reduction of 24.70\%. LOS scenarios showed a substantial average loss reduction of 42.28\%, while NLOS scenarios achieved a 13.52\% reduction. The calibration improved peak power prediction accuracy. For LOS links, the average difference between simulated and measured peak power decreased from 2.44 dB to 1.48 dB (0.96 dB improvement). For NLOS links, the average difference decreased from 8.09 dB to 7.35 dB, a 0.74 dB improvement.

\begin{figure}[!t]
\centering
\includegraphics[width=0.85\columnwidth]{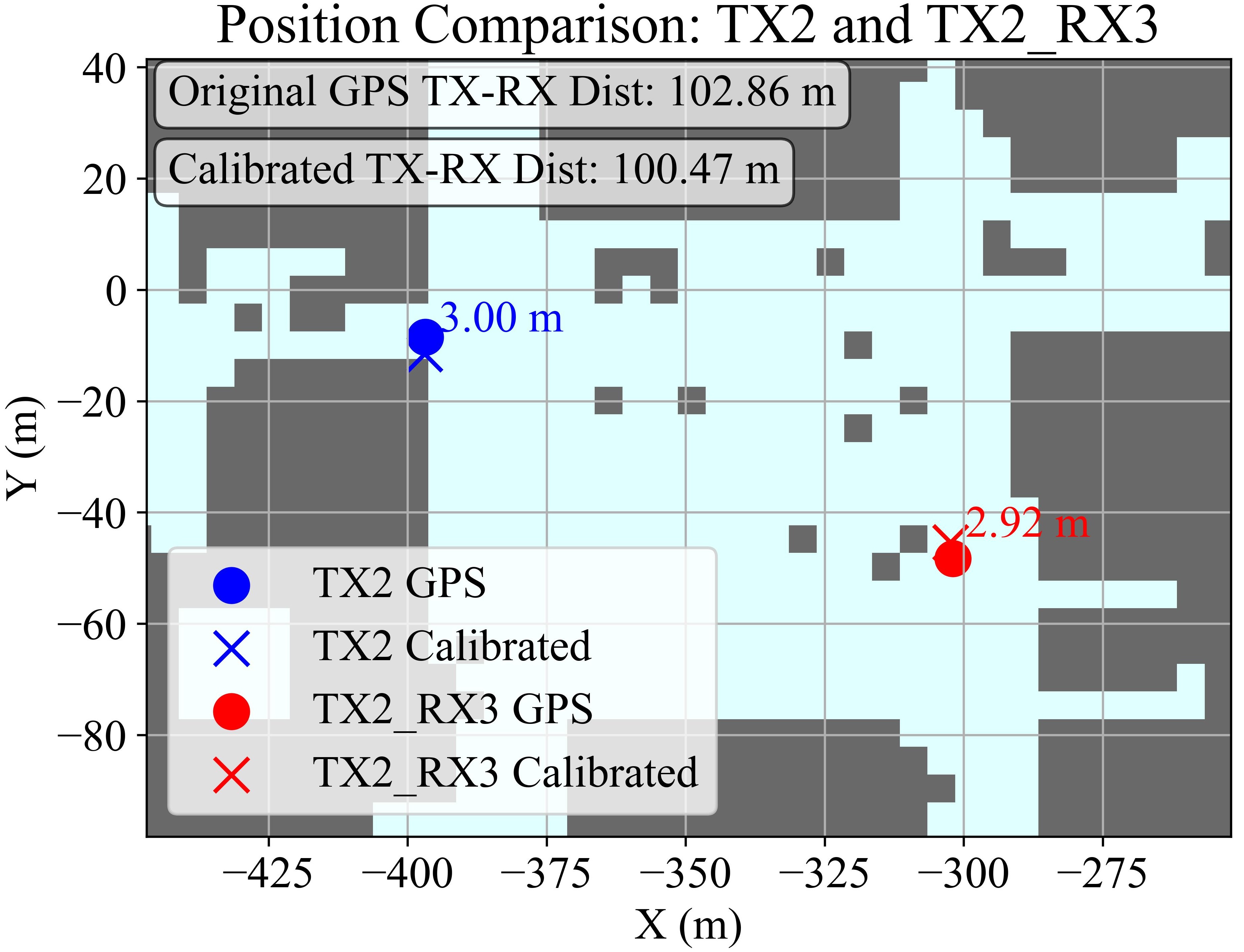}
\vspace{-5pt}
\caption{Position adjustment for TX2-RX3 pair showing original and optimized locations (TX 3 m, RX 2.92 m  adjustments).}
\label{fig:position_adjustment}
\vspace{-10pt}
\end{figure}

Fig.~\ref{fig:position_adjustment} shows the position adjustment for the TX2-RX3 pair. Original GPS-derived positions indicated a TX-RX separation of 102.86 m, whereas precise laser measurement established the ground truth distance as 100.9 m. Following calibration, the optimized T-R separation was 100.47 m. The optimization process resulted in position adjustments of 3 m for the TX and 2.92 m for the RX, highlighting significant initial GPS position uncertainty compared to the laser measurement. Across all UMi measurement pairs, the mean position adjustments were 1.69 m for TX locations and 3.09 m for RX locations.

\begin{figure}[t]
\centering
\includegraphics[width=0.9\columnwidth]{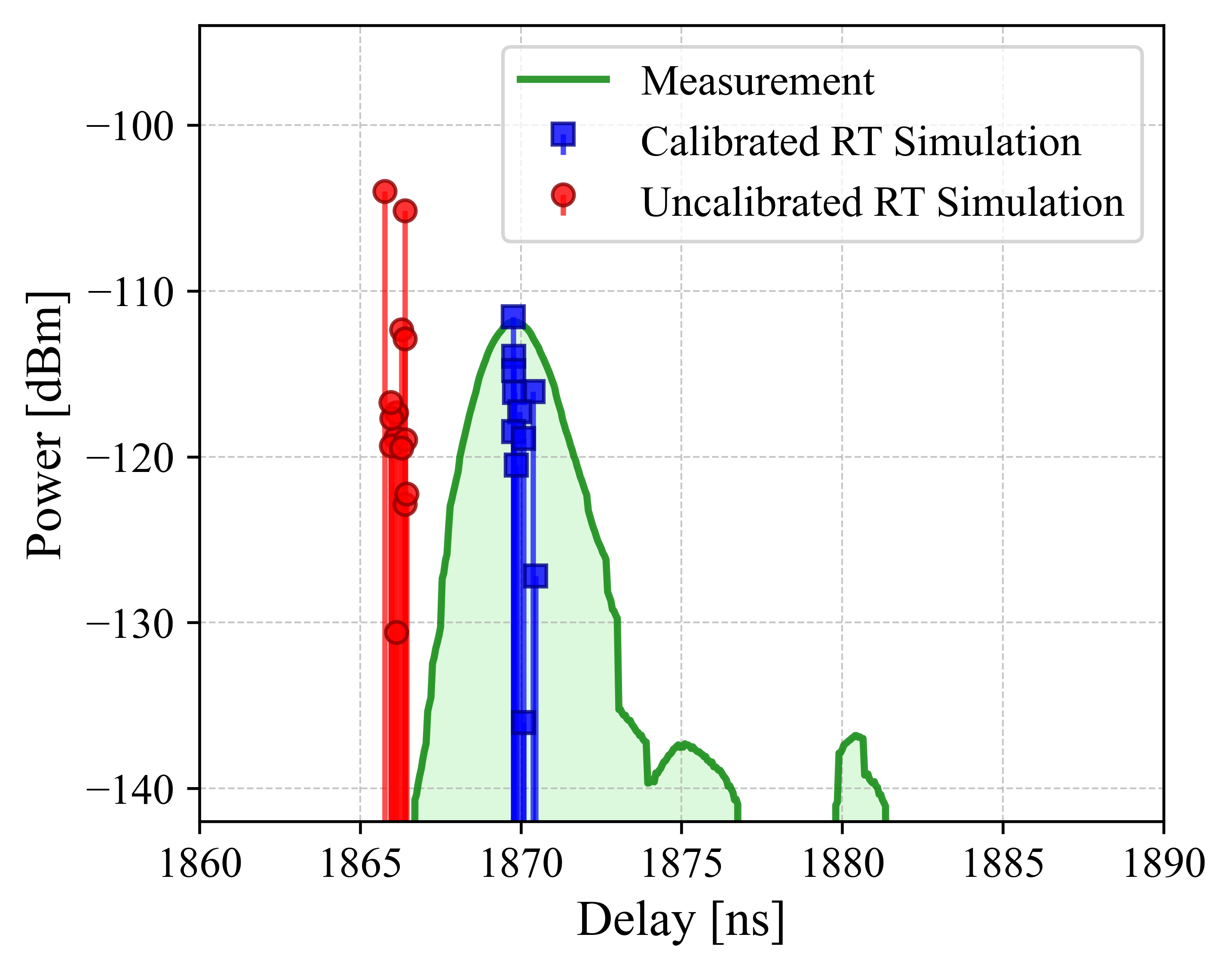}
\vspace{-5pt}
\caption{PDP comparison showing measured data versus uncalibrated location and calibrated location RT simulations for TX1-RX4 (560 m) in the UMi environment at 6.75 GHz.}
\label{fig:pdp_comparison}
\vspace{-5pt}
\end{figure}

\begin{table}[!t]
    \centering
    \caption{Calibration Performance Metrics Summary}
    \label{tab:calibration_metrics}
    \small
    \setlength{\tabcolsep}{6pt} 
    \begin{tabular}{lcccc}
    \toprule
    \multirow{2}{*}{Scenario} & Loss & \multicolumn{2}{c}{Peak Power Diff. (dB)} & $^{*}$Peak Power \\
    \cmidrule(lr){3-4}
     & Red. (\%) & Original & Calibrated & Imp. (dB)  \\
    \midrule
    LOS    & 42.28 &  2.44 & 1.48 & 0.96 \\
    NLOS   & 13.52 & 8.09 & 7.35 & 0.74 \\
    \bottomrule
    \end{tabular}
    \begin{flushleft}
    \footnotesize{$^{*}$Peak Power Imp. represents the improvement (reduction) in the absolute difference between simulated and measured peak power, comparing original RT simulation results to those after location calibration.}
    \end{flushleft}
        \vspace{-20pt}
\end{table}

NLOS scenarios generally require larger position adjustments compared to LOS scenarios. Such adjustments are expected due to the increased complexity of multipath propagation and higher sensitivity to geometry when the direct path is obstructed. While the percentage loss reduction was higher for LOS cases (42.28\% vs. 13.52\% for NLOS, see Table \ref{tab:calibration_metrics}), the calibration process addresses significant errors inherent in challenging NLOS environments. 


Fig.~\ref{fig:pdp_comparison} illustrates the impact of location calibration on the PDP for the TX1-RX4 link in the UMi environment at 6.75 GHz, characterized by a T-R separation of approximately 560 m (see Fig. 1 in \cite{Shakya2025icc}). The RT simulation using uncalibrated locations overestimates the received power by approximately 8 dB and exhibits a temporal mismatch of about 4 ns relative to the measured PDP. In contrast, the RT simulation employing calibrated locations provides a significantly improved estimation of both power and delay. The delay mismatch is reduced to within 0.1 ns, and the power difference is within 0.5 dB, demonstrating a much closer alignment with the measured data.

Matching RT-simulated PDPs with measured PDPs~\cite{Shakya2024ojcom} enables accurate TX/RX location estimation. Our location calibration method, utilizing extensive historical measurement datasets from NYU~\cite{rappaport2013ia, maccartney2015ia, Shakya2024ojcom, Ju2024twc, Xing2021_Inicl} and other institutions, provides a foundation for subsequent calibration of material properties and other RT parameters. The calibration method is easily applicable for channel measurements across frequencies (upper mid-band to sub-THz) and environments.

%% file: sections/conclusion.tex
\section{Conclusion}
\label{sec:conclusion}
Location mismatch between measurements and RT simulations challenges site-specific channel prediction, particularly at upper mid-band, mmWave, and higher frequencies, where centimeter-level position errors degrade accuracy. This paper presents a systematic multi-stage TX-RX location calibration method using a multi-component loss function through PDP alignment optimization, achieving average loss reductions of 42.28\% (LOS) and 13.52\% (NLOS). The approach reduced average peak power prediction error by 0.96~dB (LOS) and 0.74~dB (NLOS). 
While validation focused on UMi scenarios at 6.75~GHz and 16.95~GHz, the proposed location calibration framework enables more accurate utilization of past measurement datasets from NYU WIRELESS and other institutions for future RT validation and calibration. The precise spatial alignment is fundamental for developing the efficient beam management techniques and reliable network deployment strategies required by 5G-Advanced and 6G systems.